\documentclass[preprint,nofootinbib,floats,aps,showpacs,showkeys]{revtex4}
\usepackage{graphicx}
\usepackage{epsfig}
\usepackage{bm}
\usepackage{longtable}
\newcommand{\ie}{\emph{i.e.}}
\def\bit{\begin{itemize}}
\def\eit{\end{itemize}}
\def\bnu{\begin{enumerate}}
\def\enu{\end{enumerate}}

\def\nn{\nonumber }

\def\M {{{\cal M}}}
\def\x{\times}
\def\Ket#1{||#1 \rangle}
\def\Bra#1{\langle #1||}

\def\ie{{\em i.e., }}

\def\nn{\nonumber }
\def\be{\begin{equation}}
\def\ee{\end{equation}}
\def\br{\begin{eqnarray}}
\def\er{\end{eqnarray}}
\def\brn{\begin{eqnarray*}}
\def\ern{\end{eqnarray*}}
\def\etc{ {\it etc}}

\def\pb {{\bf p}}
\def\Pb{ {\bf P}}
\def\lb {{\bf l}}
\def\Lb{ {\bf L}}
\def\rb {{\bf r}}
\def\Rb{ {\bf R}}
\def\e {{\epsilon}}
\def\mbl{\mbox{\boldmath$\lambda$}}
\def\bra#1{\langle #1|}
\def\ket#1{|#1 \rangle}
\def\rf#1{{(\ref{#1})}}

\def\sixj#1#2#3#4#5#6{\left\{\negthinspace\begin{array}{ccc}
#1&#2&#3\\#4&#5&#6\end{array}\right\}}

\def\go{\rightarrow  }
\def\etal{{\it et al.}}
\def\F {{{\cal F}}}
\def\fot{\frac{1}{2}}
\def\ve{\varepsilon}
\newcommand{\rn}{\mathrm{n}}
\newcommand{\rP}{\mathrm{P}}

\begin{document}

\title{Kinetic energy sum  spectra  in
nonmesonic weak  decay of hypernuclei}

\author{Cesar Barbero$^{1,4}$}

\author{Alfredo P. Gale\~ao$^2$}

\author{Mahir S. Hussein$^{3,5}$}

\author{Francisco Krmpoti\'c$^{3,4,6}$}

\affiliation{$^1$
Facultad de Ciencias Exactas, Departamento de F\'isica, Universidad Nacional de
La Plata, 1900 La Plata, Argentina}
\affiliation{$^2$
Instituto de F\'{\i}sica Te\'orica,
Universidade Estadual Paulista, \\
Rua Pamplona 145, 01405-900 S\~ao Paulo, SP, Brazil}
\affiliation{$^3$Departamento de F\'isica Matem\'atica, Instituto
de F\'isica da Universidade de S\~ao Paulo,
Caixa Postal 66318, 05315-970 S\~ao Paulo, SP, Brazil}
\affiliation{$^4$Instituto de F\'isica La Plata, CONICET, 1900 La
Plata, Argentina}
\affiliation{$^5$ Max-Planck-Institut f\"ur Physik komplexer Systeme,
N\"othnitzer Stra{\ss}e 38, D-01187 Dresden, Germany}
\affiliation{$^6$Facultad de Ciencias Astron\'omicas y
Geof\'isicas, Universidad Nacional de La Plata, 1900 La Plata,
Argentina}

\date{\today}

\begin{abstract}

We evaluate  the  coincidence spectra in the
nonmesonic weak decay (NMWD) $\Lambda N\go nN$ of $\Lambda$ hypernuclei
 $^{4}_\Lambda$He, $^{5}_\Lambda$He, $^{12}_\Lambda$C,
$^{16}_\Lambda$O, and $^{28}_\Lambda$Si,
as a function of the  sum of kinetic energies $E_{nN}=E_n+E_N$ for
$N=n,p$. The  strangeness-changing transition potential is
described by the one-meson-exchange model, with  commonly used
parameterization. Two versions of the  Independent-Particle-Shell-Model (IPSM) 
are employed to account for the nuclear structure of  the final
residual nuclei. They are: (a) IPSM-a, where  no correlation, except for the 
Pauli principle, is  taken into account, and (b)
IPSM-b, where the highly excited  hole states are considered to be 
quasi-stationary and are described by  Breit-Wigner
distributions, whose widths are estimated from the
experimental data. All  $np$ and $nn$ spectra exhibit a series of
peaks in the energy interval $110$ MeV $<E_{nN}<170$ MeV, one
for each occupied shell-model state. Within the IPSM-a, and because
of the recoil effect, each peak covers an energy interval
proportional to $A^{-1} $,  going from $\cong 4$ MeV for
$^{28}_\Lambda$Si to  $\cong 40$ MeV for $^{4}_\Lambda$He. Such a
description could be  pretty fair for the light $^{4}_\Lambda$He and
$^{5}_\Lambda$He hypernuclei. For the remaining, heavier,  hypernuclei
it is very important, however, to consider as well
the spreading in strength of the deep-hole states, 
and  bring into play  the IPSM-b approach. Notwithstanding
the nuclear model that is employed the results depend only very weakly on the
details of the dynamics involved in the decay process proper.
We propose that the IPSM is the appropriate lowest-order
approximation for the theoretical calculations of the of kinetic energy sum  
spectra  in the NMWD. It is in comparison to this picture that one should 
appraise the effects of the  final state interactions and of the 
two-nucleon-induced decay mode.

\end{abstract}

\pacs{21.80.+a, 13.75.Ev, 21.60.Cs, 21.10.Pc}

\keywords{nonmesonic hypernuclear decay; energy spectra; strength functions;
one meson exchange model}

\maketitle

\section{Introduction \label{Int}}

Because of the difficulty in detecting neutrons, up to a few years
ago only the high energy proton spectra had been measured in the
two-body $\Lambda N\go nN$ nonmesonic weak decay   (NMWD) of
$\Lambda$ hypernuclei. The proton-induced transition rates
$\Gamma_p\equiv\Gamma(\Lambda p\go np)$ in $^{5}_\Lambda$He,
$^{11}_\Lambda$Li, and $^{12}_\Lambda$C have been determined in
this way~\cite{Sz91,No95,Ha02}. The corresponding neutron-induced
transition rates $\Gamma_n\equiv\Gamma(\Lambda n\go nn)$ were
estimated through the comparison of the measured proton spectrum
with that  of the intranuclear cascade (INC) calculation where the
$\Gamma_n/\Gamma_p$ ratio is treated as free parameter. This
procedure signalized  very large  experimental $n/p$ ratios
($\cong 0.9-2.0$)~\cite{Sz91,No95,Ha02}, in comparison with
theoretical estimates ($\cong 0.3-0.5$) obtained within the
One-Meson-Exchange Potential (OMEP) ~\cite{Pa01,It02,
Ba02,Kr03,Ba03,Ba04,Ba05,Ga07}. (See, for instance, \cite[Table
3]{Kr03}.). In spite of  large uncertainties involved in  the
indirect evaluation of neutrons, this discrepancy between data
and  theory was considered to be a serious puzzle in the NMWD.

Yet, quite recently, the above scenario has drastically changed  due to
the very significant advances in our knowledge (mainly  experimental)
on both  neutron and proton spectra. These advances are:

\bnu

\item   In the experiment E$369$ were measured high-quality neutron spectra in
the decays of $^{12}_\Lambda$C and $^{89}_\Lambda$Y, which made possible to
compare them  directly  with the corresponding proton  spectra, yielding the
result $\Gamma_n/\Gamma_p=(0.45-0.51)\pm 0.15$ \cite{Ki03}.

\item Okada \etal\ \cite{Ok04} have simultaneously measured the energy spectra
of neutrons and protons in $^{5}_\Lambda$He and $^{12}_\Lambda$C at a high
energy threshold ($60$ MeV), from where it was inferred that for both
hypernuclei $\Gamma_n/\Gamma_p\cong 0.5$.

\item  Garbarino,  Parre\~no and  Ramos~\cite{Ga03,Ga04,Ga05} have called
attention on the fact  that ``correlation observables  permit a {\it cleaner}
extraction of $\Gamma_n/\Gamma_p$ from data than single-nucleon observables",
which has stimulated several experimental searches~\cite{Ok05,Ou05,Ka06,Ki06}.
They have also done a theoretical evaluation of the pair distributions for:
a) the sum of the $nN$  kinetic energies $E=E_n+E_N$, $S_{nN}(E)$, and b) the
opening angle $\theta$, $S_{nN}(\cos\theta)$, in the decays of $^{5}_\Lambda$He 
and $^{12}_\Lambda$C. The number of detected $nN$ pairs ${\rm N}_{nN}$ is 
proportional to 
$ \Gamma_{N}=\int S_{nN}(E)dE=\int S_{nN}(\cos\theta)d\cos\theta$, and
therefore ${\rm N}_{nn}/{\rm N}_{np}\cong \Gamma_{n}/\Gamma_{p}$ (see below).
The primary  weak decay was described by the OMEP dynamics within the framework 
of a shell model, while an INC code was used to take  into account the strong 
final state interactions (FSI) involving the primary nucleons and
those in the residual nucleus, including the possibility of emission of
secondary particles.
They conclude that the datum ${\rm N}_{nn}/{\rm N}_{np} = 0.44 \pm 0.11$ 
obtained in KEK-E462 \cite{Ou05} for $^{5}_\Lambda {\rm He}$ is compatible 
with an $n/p$ ratio of  $0.39 \pm 0.11$ for this hypernucleus if the 
two-nucleon induced mode $\Lambda NN\go nNN$ is neglected, or an even lower 
value if it is included. Similarly, together with Bauer \cite{Ba06}, they found
$\Gamma_n/\Gamma_p(^{12}_\Lambda {\rm C})=0.46\pm 0.09$ when the two-nucleon
induced mode is neglected and a slightly lower value when it is included.
The shell-model used in  Refs.~\cite{Ga03,Ga04,Ga05} is substituted in the 
last work  by a nuclear matter formalism  extended to finite nuclei via the 
local density approximation, while the Monte Carlo INC model is retained to 
account for the FSI.

\item Quite recently Bauer \cite{Bau07} has given a step forward with his 
nuclear matter formalism, describing microscopically both   the weak decay 
mechanism and the FSI, confirming in this way that the latter lead to somewhat 
lower value for the ratio $\Gamma_n/\Gamma_p$ in $^{12}_\Lambda$C.

\item Several coincidence emission measurements of the above mentioned spectra
have been performed~\cite{Ok05,Ou05,Ka06,Ki06,Pa07,Bh07}, from 
which were extracted the  new experimental results,  
$\Gamma_n/\Gamma_p(^{5}_\Lambda {\rm He})=0.45\pm 0.11\pm 0.03$ and
$\Gamma_n/\Gamma_p(^{12}_\Lambda {\rm C})=0.40\pm 0.09$, 
that point towards the solution of the longstanding $\Gamma_n/\Gamma_p$ puzzle.%
\footnote{We  note that the relationship
$\Gamma_n/\Gamma_p(^{5}_\Lambda {\rm He}) \sim
\Gamma_n/\Gamma_p(^{12}_\Lambda {\rm C})$ could have  a very simple explanation,
similar to that given in Ref. \cite{Ba07} for the asymmetry parameter:
$a_\Lambda(^{5}_\Lambda {\rm He})\sim a_\Lambda (^{12}_\Lambda {\rm C})$.}

\enu

Reasoning within the two-body kinematics for the one-nucleon induced  NMWD,
it is expected that $S_{nN}(E)$  should exhibit a narrow peak
at the two-particle energy $E$ close to the decay Q-value, while
$S_{nN}(\cos\theta)$ should be restrained  within  the back-to-back angle
$\theta\cong \pi$. Thus, all experimentally observed  deviations from such
spectral shapes are very frequently attributed to the FSI and/or to the
two-nucleon induced processes
~\cite{Ok05,Ou05,Ka06,Ki06,Pa07,Bh07}.%
\footnote{In Refs.~\cite{Ga03,Ga04} it is said that the $np$ energy spectra of
the three-body proton-induced decay $ ^{5}_\Lambda$He $\go ^{3}$H $+n+p$ should
exhibit a narrow peak close to its Q-value of $153$ MeV,
which  is only valid when the recoil effect is neglected.}
However, this is no longer the case  when the recoil of the residual nucleus is
taken into account, which makes the kinematics to be of a three-body type.
Moreover, when the shell model structure is also taken into account the energy
spectra will have a bump at each single particle state, the width of which
will depend on the magnitude of  both  the recoil and the spreading in strength 
of the hole states in the inner shells.
Although  the detailed structure and fragmentation
of hole states are still not well known, the
exclusive knockout reactions provide a wealth of
information on the structure of single-nucleon states
of nuclei. Excitation energies  and widths of
proton-hole states were systematically measured with
quasifree $(p, 2p)$ and $(e, e' p)$ reactions, which
revealed the existence of inner orbital shells in 
nuclei~\cite{Ja73,Fr84,Be85,Le94,Ya96,Ya01,Yo03,Ya04,Ko06}.

The aim of the present work is to discuss quantitatively the interplay  between 
the recoil effect and the nuclear shell structure in the kinetic energy sum 
spectra of NMWD.
The paper is organized as follows. In Section \ref{Kin} we
discuss the calculation of these spectra within the  
Independent-Particle-Shell-Model (IPSM). In Section \ref{Num} we 
exhibit the numerical results for  $^{4}_\Lambda$He, $^{5}_\Lambda$He, 
$^{12}_\Lambda$C, $^{16}_\Lambda$O, and $^{28}_\Lambda$Si hypernuclei.
In Section \ref{Gen} we discuss these results and their connection with the 
experimental data.
Finally,  in Section \ref{Con},  we present several concluding remarks.

\section{ Kinetic energy sum spectrum \label{Kin}}

For the purpose of completeness and clarity,  we will first redo
the calculation of the  decay rate $\Gamma_N$ using Fermi's
golden  rule.
The novelties  here,  in comparison with our previous works
Refs.~\cite{Ba02,Kr03,Ba03}, are the recoil effect and the spreading of
the single-particle configurations. As will be shown in Section \ref{Gen}, 
their role is of minor importance in the evaluation of the integrated  
transition rates $\Gamma_N$, as well as on the ratio $\Gamma_n/\Gamma_p$, but 
they are crucial for a correct description of the energy distribution of the 
transition strength.
In fact, a single-particle state $\ket{j_N}$ that is  deeply bound in the 
hypernucleus,  after the NMWD can become a highly excited hole-state 
$\ket{j^{-1}_N}$ in the continuum of the residual nucleus. 
There  it suddenly mixes up with more complicated configurations 
(2h1p, 3h2p, \dots excitations, collective states, \etc.) 
spreading its strength in a relatively wide energy interval~\cite{Ma85}.%
\footnote{One should keep in mind that the mean life a $\Lambda$ hyperon is 
$\tau_\Lambda= 2.63 \times 10^{-10}$ s, while the strong interaction times are 
of the order of $10^{-21}$ s.}
This happens, for instance, with the $1s_{1/2}$ orbital in $_\Lambda^{12}$C,
that is separated  from the $1p_{3/2}$ state by approximately $23$ MeV, which 
is enough to break the $10$ particle system, where the energy of the last 
excited state amounts to $\sim 16.5$ MeV.

The NMWD rate  of a hypernucleus (in its ground state with spin $J_I$ and 
energy $E_{J_I}$) to residual nuclei (in the several allowed states with spins 
$J_F$ and energies $E_{\alpha_NJ_F}$) and two free nucleons $nN$ (with  total 
spin $S$ and total kinetic  energy $E_{nN}=E_n+E_{N}$), reads
\begin{eqnarray}
\Gamma_N &=& 2\pi \sum_{SM_S\alpha_NJ_FM_F} 
\int |\bra{\pb_n\pb_N SM_S;\alpha_NJ_FM_F}V\ket{J_IM_I}|^2 
\nonumber \\
&\times& \delta(\Delta_{\alpha_NJ_F}-E_r-E_{nN}) 
\frac{d{\bf p}_n}{(2\pi)^3}\frac{d{\bf p}_N}{(2\pi)^3}.
\label{1}
\end{eqnarray}
Here, $V$ is  the hypernuclear nonmesonic weak transition potential, 
and the wave functions for the kets $\ket{\pb_n\pb_N SM_S;\alpha_NJ_FM_F}$ and 
$\ket{J_IM_I}$ are assumed to be antisymmetrized  and normalized.
The label ${\alpha_N}$ stands for different final states  with the same spin 
$J_F$, $E_{r}$ is the recoil energy of the residual nucleus, and
\begin{equation}
\Delta_{\alpha_NJ_F}=\Delta+E_{J_I}-E_{\alpha_NJ_F}, 
\hspace{1cm} \mbox{with} \hspace{1cm}
\Delta=M_\Lambda-M=176~~ \mbox {MeV},
\label{2}
\end{equation}
is the liberated energy. The two emitted nucleons are described by plane waves, 
and initial and final short range correlations are included phenomenologically 
at a simple Jastrow-like level.

It is convenient to perform a transformation to the  relative  and
c.m. momenta ($\pb=\fot(\pb_n-\pb_N $),  $\Pb=\pb_n+\pb_N$), coordinates
($\rb=\rb_n-\rb_N$,  $\Rb=\fot(\rb_n+\rb_N)$) and orbital angular momenta
$\lb$ and $\Lb$,
and to express the energy conservation as
\be
E_{nN}+E_r-\Delta_{\alpha_NJ_F}=\e_p+\e_P-\Delta_{\alpha_NJ_F}=0,
\label{3}\ee
where
\be
\e_p=\frac{p^2}{M},
\hspace{1cm}E_r=\frac{P^2}{2M(A-2)},
\hspace{1cm}\e_P=\frac{P^2}{4M}\frac{A}{A-2}=\frac{A}{2}E_r,
\label{4}\ee
are, respectively, the energies of  the relative motion of the outgoing pair, 
of the recoil, and of the total c.m. motion (including the recoil).
Following step by step the analytical developments
done in Ref.~\cite{Ba02},   the transition rate can be can expressed as
 \br
\Gamma_{N}&=&\int_{0}^{\Delta}\frac{d\Gamma_{N}}{d\e_P} d\e_P
\label{5}\er
where we have defined  (see~\cite[Eqs. (2.13) and (2.14)]{Ba02})
\br
\frac{d\Gamma_{N}}{d\e_P}&=&
\frac{16M^3}{\pi}\left(\frac{A-2}{A}\right)^{3/2}
\hat{J}_I^{-2}\sum_{S\lambda lLTJ\alpha_NJ_F}
\sqrt{\e_P(\Delta_{\alpha_NJ_F}-\e_P)}
\nn\\
&\x&\left|\sum_{j_N} \M(pPlL\lambda SJT;{j_N})
\Bra{J_I}\left( a_{j_N}^\dag a_{j_\Lambda}^\dag\right)_{J}
\Ket{\alpha_NJ_F}\right|^2,
\label{6}\er
it being understood that the square root is to be replaced by zero whenever its 
argument is negative. The angular momentum couplings ${\bf l}+{\bf L}={\mbl}$ 
and ${\mbl}+{\bf S}={\bf J}$ have been carried out, $\hat{J}\equiv\sqrt{2J+1}$,
and $A=Z+N+1$ is the total number of baryons.

It is self-evident that for $A\go \infty$ one obtains the same
result as in Refs. \cite{Ba02,Kr03,Ba03}. It is also worth noting
that the overall outcome of the recoil on $\Gamma_N$ is very small,
mostly because the effect of  the factor
$\left(\frac{A-2}{A}\right)^{3/2}$ in Eq. \rf{6} is, to a great
extent, cancelled by the effect of  the factor
$\left(\frac{A}{A-2}\right)^{3/2}$ originating from
$\sqrt{\e_P(\Delta_{\alpha_NJ_F}-\e_P)} d\epsilon_P$. This is the
reason why we have not included the recoil previously.

The  spectrum  of $\Gamma_N$ as a function of $E_{nN}$ is now
easily obtained from Eq. \rf{6} by means of the relation
\be
E_{nN}=\Delta_{\alpha_NJ_F}-\frac{2}{A}\e_P,
\label{7}\ee
as follows from \rf{3} and \rf{4}. Calling $E \equiv E_{nN}$, one gets
\begin{equation}
\Gamma_N = \int_0^\Delta S_{nN}(E)\, dE 
\label{extra8}
\end{equation}
with
\br
S_{nN}(E) &=&\frac{4M^3}{\pi}\sqrt{A(A-2)^3}
\hat{J}_I^{-2}\sum_{S\lambda lLTJ\alpha_NJ_F}
\sqrt{(\Delta_{\alpha_NJ_F}-E)(E-\Delta_{\alpha_NJ_F}')}
\nn\\
&\x&\left|\sum_{j_N} \M(pPlL\lambda SJT;{j_N})
\Bra{J_I}\left( a_{j_N}^\dag a_{j_\Lambda}^\dag\right)_{J}
\Ket{\alpha_NJ_F}\right|^2,
\label{8}\er
where 
\br
p&=&\sqrt{\frac{MA}{2}\left(E-\Delta_{\alpha_NJ_F}'\right)},
\nn\\
{P}&=&\sqrt{2M(A-2)(\Delta_{\alpha_NJ_F}-E)},
\label{9}\er
\be
\Delta_{\alpha_NJ_F}'=\Delta_{\alpha_NJ_F}\frac{A-2}{A},
\label{10}\ee
and the condition
\be
\Delta_{\alpha_NJ_F}'\le E\le \Delta_{\alpha_NJ_F},
\label{11}\ee
has to be fulfilled for each contribution.

As previously \cite{Ba02,Kr03,Ba03}, it will be assumed that the hyperon 
in the state ${j_\Lambda}$, with  single-particle energy
$\e_{j_\Lambda}$,  is  weakly coupled to the $A-1$ core, 
with spin ${J_C}$ and energy $E_C=E_{J_I}-\e_{j_\Lambda}$. 
Then the initial state is $\ket{J_I}\equiv\ket{(J_Cj_\Lambda)J_I}$, and the 
spectroscopic amplitude 
$\Bra{J_I}\left( a_{j_N}^\dag a_{j_\Lambda}^\dag\right)_{J}\Ket{\alpha_NJ_F}$ 
can be rewritten as
\br
\Bra{J_I}\left( a_{j_N}^\dag
a_{j_\Lambda }^\dag\right)_{J}\Ket{\alpha_NJ_F}
&=&(-)^{J_F+J+J_I}\hat{J}\hat{J}_I
\sixj{J_C}{J_I}{j_\Lambda}{J}{j_N}{J_F} \Bra{J_C}a_{j_N}^\dag\Ket{\alpha_NJ_F}.
\label{12}\er

The following two approaches  for the final states $\ket{\alpha_NJ_F}$ 
will be examined within the IPSM.

\subsection{IPSM-a}

Here, we completely ignore the residual interaction and, consequently, the only 
states $\ket{\alpha_NJ_F}$ giving a nonzero result in Eq. \rf{12} and therefore 
contributing to Eq. \rf{8} are those obtained by the weak coupling,  and
properly antisymmetrizing,  of the one  hole (1h) states $\ket{j^{-1}_N}$
to the core ground-state $\ket{J_C}$.  That is, recalling that we completely 
ignore the residual interaction in this approximation,
\begin{equation}
\ket{\alpha_N J_F} \mapsto \ket{j_NJ_F} \equiv \ket{(J_C,j^{-1}_N)J_F}\,,
\quad \mathrm{and} \quad
E_{\alpha_N J_F} \mapsto E_{j_N} \equiv  E_C - \epsilon_{j_N} \,,
\label{13}\end{equation}
where $\epsilon_{j_N}$ is the single-particle energy of state $j_N$.
As an illustration, in the case of $^{28}_\Lambda$Si the model space contains
four single-particle states, both for protons and for neutrons
($\rn_p=\rn_n=4$), namely, $1s_{1/2}$, $1p_{3/2}$, $1p_{1/2}$ and $1d_{5/2}$.
Thus,  for $\ket{J_C}=\ket{1d_{5/2}n^{-1}}$,
the final states \rf{13} are constructed by adding two holes in the 
$^{28}$Si nucleus, and read:
\br
\begin{array}{ll}
\underline{_\Lambda^{28}{\rm Si}\rightarrow nn+{^{26}{\rm Si}}}&
\underline{_\Lambda^{28}{\rm Si}\rightarrow np+ {^{26}{\rm Al}}}\\\\
\ket{(1d_{5/2}n^{-1})^2;0,2,4} & 
\ket{(1d_{5/2}n^{-1}1d_{5/2}p^{-1});0,1,2,3,4,5}\\
\ket{1d_{5/2}n^{-1}1s_{1/2}n^{-1};2,3} &  
\ket{1d_{5/2}n^{-1}1s_{1/2}p^{-1};2,3}\\
\ket{1d_{5/2}n^{-1}1p_{1/2}n^{-1};2,3} &  
\ket{1d_{5/2}n^{-1}1p_{1/2}p^{-1};2,3}\\
\ket{1d_{5/2}n^{-1}1p_{3/2}n^{-1};1,2,3,4} &  
\ket{1d_{5/2}n^{-1}1p_{3/2}p^{-1};1,2,3,4}.
\end{array}
\label{14}\er

The summation on ${J_F}$ in  \rf{8} can be performed
for each single-particle state $j_N$, as done  
in~\cite[Eqs. (11), (12), (13)]{Kr03}.  One gets
\br
{S}_{nN}(E) &=&\frac{4M^3}{\pi}\sqrt{A(A-2)^3}\sum_{j_N}
\sqrt{(\Delta_{j_N}-E)(E-\Delta'_{j_N})}\F_{j_N}(pP),
\label{15}\er
where
\br
\F_{j_N}(pP)&=&\sum_{ J=|j_N-1/2|}
^{J=j_N+1/2}F^J_{j_N}\sum_{SlL\lambda T} \M^2(pPlL\lambda SJT;j_N),
\label{16}\er
with the spectroscopic factors $F^J_{j_N}$ exhibited in~\cite[Table 1]{Kr03}.
The maximum and minimum liberated energies are, respectively,
\be
\Delta_{\alpha_N J_F} \mapsto 
\Delta_{j_N} = \Delta + \epsilon_{j_\Lambda} + \epsilon_{j_N}
\label{17}\ee
and
\be
\Delta'_{j_N}=\Delta_{j_N}\frac{A-2}{A},
\label{18}\ee
and the momenta are given by
\br
p&=&\sqrt{\frac{MA}{2}\left(E-\Delta'_{j_N}\right)}
\hspace{0.5cm} \mathrm{and} \hspace{0.5cm} 
P=\sqrt{2M(A-2)(\Delta_{j_N}-E)}.
\label{19}\er

\subsection{IPSM-b}

Formally, one starts from the unperturbed basis $\ket{i_N J_F}_0$ with 
$i_N=1,2,\dots \rn_N,\rn_N+1,\dots $, where for $i_N\le \rn_N$ we have the same 
simple doorway states $\ket{j_NJ_F}$ in Eq. \rf{13}
(listed in Eq. \rf{14} for $^{28}_\Lambda$Si), while for $i_N \ge \rn_N+1$ we 
have more complicated bound configurations 
(such as $3h1p$, $4h2p$, \dots in the case of $^{28}_\Lambda$Si) as
well as those including unbound single-particle states in the continuum.
As in Ref.~\cite{Ma85}, the  perturbed  eigenkets  $\ket{\alpha_N J_F}$ and  
eigenvalues $E_{\alpha_N J_F}$ are obtained by diagonalizing the matrix 
$_0\bra{i_N J_F}H\ket{i'_N J_F}_0 $ of the exact Hamiltonian $H$:
\begin{equation}
\bra{\alpha_N J_F}H\ket{\alpha'_N J_F } = 
E_{\alpha_N J_F} \, \delta_{\alpha_N \alpha'_N}
\label{20}
\end{equation}
with
\begin{eqnarray}
\ket{\alpha_N J_F} &=& \sum_{i_N=1}^\infty C_{i_N}^{\alpha_NJ_F} \ket{i_N J_F}_0
\nonumber \\
&=& \sum_{j_N} C_{j_N}^{\alpha_NJ_F} \ket{j_NJ_F}
+ \sum_{i_N=\rn_N+1}^\infty C_{i_N}^{\alpha_NJ_F} \ket{i_N J_F}_0 \,.
\label{21}
\end{eqnarray}
It is easy to see that only the ket $\ket{j_NJ_F}$ in the expansion \rf{21} 
will contribute to the matrix element 
$\bra{J_C}|a^\dagger_{j_N}|\ket{\alpha_N J_F}$ in Eq. \rf{12}. 
Therefore Eq. \rf{8} takes the form
\begin{eqnarray}
S_{nN}(E) &=& \frac{4M^3}{\pi} \sqrt{A(A-2)^3}
\nonumber \\ &\times&
\sum_{j_N\alpha_NJ_F} |C_{j_N}^{\alpha_NJ_F}|^2
\sqrt{(\Delta_{\alpha_NJ_F}-E)(E-\Delta'_{\alpha_NJ_F})}
\mathcal{F}(p,P;j_NJ_F) \,,
\label{22}
\end{eqnarray}
where
\begin{eqnarray}
\mathcal{F}(p,P;j_NJ_F) &=& \hat{J}_I^{-2} \sum_{lL\lambda SJT}
\left| \mathcal{M}(pPlL\lambda SJT;j_N)
\bra{J_I}|\left(a^\dagger_{j_N} a^\dagger_{j_\Lambda}\right)_J|\ket{j_NJ_F} 
\right|^2 \,.
\label{23}
\end{eqnarray}

To evaluate the amplitudes $C_{j_N}^{\alpha_NJ_F}$  one would have to choose  
the appropriate Hamiltonian $H$ and the unperturbed basis $\ket{i_N J_F}_0$, 
and solve the eigenvalue problem \rf{20}.
We will not do this here. Instead, we will make a phenomenological 
estimate. First, because of the high density of states, we will convert the 
discrete energies $\Delta_{\alpha_NJ_F}$ into the  continuous variable
$\ve$, and the discrete  sum on ${\alpha_N}$ into an integral on $\ve$, \ie
\be
\Delta_{\alpha_NJ_F} \to \varepsilon \,, \quad
\sum_{\alpha_NJ_F} |C_{j_N}^{\alpha_NJ_F}|^2  \to
\sum_{J_F} \int_{-\infty}^{\infty} |C_{j_NJ_F}(\varepsilon)|^2 
\rho_{J_F}(\varepsilon) d\varepsilon  \,,
\label{24}\ee
where $\rho_{J_F}(\varepsilon)$ is the density of perturbed states with 
angular momentum $J_F$. In this way the spectrum \rf{22} becomes
\begin{eqnarray}
S_{nN}(E) &=& \frac{4M^3}{\pi} \sqrt{A(A-2)^3}
\nonumber \\ &\times&
\sum_{j_NJ_F} \int_{-\infty}^{\infty} \rP_{j_NJ_F}(\varepsilon)
\sqrt{(\varepsilon-E)(E-\varepsilon')}
\mathcal{F}(p,P;j_NJ_F)  d\varepsilon \,,
\label{25}
\end{eqnarray}
where
\begin{equation}
\rP_{j_NJ_F}(\varepsilon) = |C_{j_NJ_F}(\varepsilon)|^2 \rho_{J_F}(\varepsilon)
\label{26}
\end{equation}
is called the \emph{strength function} \cite{Ma85,Fraz96,Sarg00} and represents 
the probability of finding the configuration
$\ket{j_NJ_F} \equiv \ket{(J_C,j_N^{-1})J_F}$ per unit energy interval. 
Moreover, 
\br
p&=&\sqrt{\frac{MA}{2}\left(E-\ve'\right)},
\nn\\
{P}&=&\sqrt{2M(A-2)(\ve-E)},
\label{27}\er
\be
\ve'=\ve\frac{A-2}{A},
\label{28}\ee
and the condition
\be
\ve'\le E\le \ve,
\label{29}\ee
has to be fulfilled throughout the $\varepsilon$ integration. It is convenient 
to introduce the averaged strength function
\begin{equation}
\rP_{j_N}(\varepsilon) = \frac{1}{\dim(j_NJ_C)}
\sum_{J_F=|J_C-j_N|}^{J_C+j_N} \rP_{j_NJ_F}(\varepsilon) \,,
\label{30}
\end{equation}
where
\begin{equation}
\dim(j_NJ_C) = \left\{
\begin{array}{lcl}
2j_N+1&\mathrm{for}&j_N \le J_C \,,
\\
2J_C+1&\mathrm{for}&J_C < j_N \,.
\end{array}
\right.
\label{31}
\end{equation}
This allows to simplify Eq.~\rf{25} by making the approximation 
$\rP_{j_NJ_F}(\varepsilon) \approx \rP_{j_N}(\varepsilon)$ to get
\begin{eqnarray}
S_{nN}(E) &=& \frac{4M^3}{\pi} \sqrt{A(A-2)^3}
\nonumber \\ &\times&
\sum_{j_N} \int_{-\infty}^{\infty} \rP_{j_N}(\varepsilon)
\sqrt{(\varepsilon-E)(E-\varepsilon')}
\mathcal{F}_{j_N}(pP)  d\varepsilon \,,
\label{32}
\end{eqnarray}
where we noticed that the summation of $\mathcal{F}(p,P;j_NJ_F)$ over $J_F$ 
gives the function defined in Eq.~\rf{16}.

The IPSM-a results would be recovered if one made the further 
approximation
\be
{\rm P}_{j_N}(\ve)=\delta(\ve-\Delta_{j_N}).
\label{33}\ee
Here, in IPSM-b, the $\delta$-functions \rf{33} will be used for the 
strictly stationary states, while for the fragmented hole states we will use
Breit-Wigner distributions,
\be
{\rm P}_{j_N}(\ve)=\frac{2\gamma_{j_N}}{\pi}
\frac{1}{\gamma_{j_N}^2+4(\ve-\Delta_{j_N})^2},
\hspace{1cm}
\int^\infty_{-\infty}{\rm P}_{j_N}(\ve)d\ve=1,
\label{34}\ee
where  $\gamma_{j_N}$ are the widths of the resonance centroids at energies 
$\Delta_{j_N}$ (see~\cite[Eq.(2.11.22)]{Ma85}).

It might be important to point out that, since both strength functions 
${\rm P}_{j_N}(\ve)$ are normalized to unit, their effect on integrated 
observables like the decay rates $\Gamma_N$ is expected to be small 
even if they considerably affect the spectra.
This will be further investigated in Section \ref{Gen}.

\section{Numerical Results \label{Num}}

\begin{figure}[h]
\begin{center}
   \leavevmode
   \epsfxsize = 14cm
     \epsfysize = 9cm
\epsffile{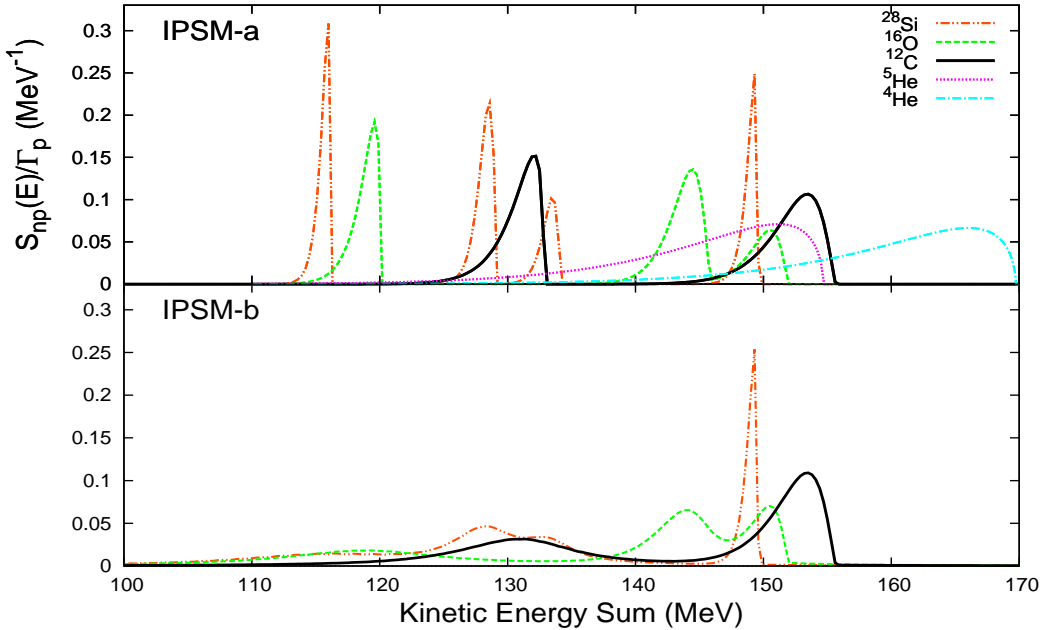}
\end{center}
\caption{\footnotesize
(Color online) Normalized energy spectra  $ S_{np}(E)/\Gamma_p$ for
$^{4}_\Lambda$He, $^{5}_\Lambda$He, $^{12}_\Lambda$C, $^{16}_\Lambda$O, and
$^{28}_\Lambda$Si hypernuclei for the full OMEP
obtained within the approaches IPSM-a (upper panel) and  IPSM-b (lower panel).
For the $s$-shell hypernuclei, only the IPSM-a approach has been used.}
\label{Fig1}\end{figure}

\begin{figure}[h]
\begin{center}
   \leavevmode
   \epsfxsize = 14cm
     \epsfysize = 9cm
\epsffile{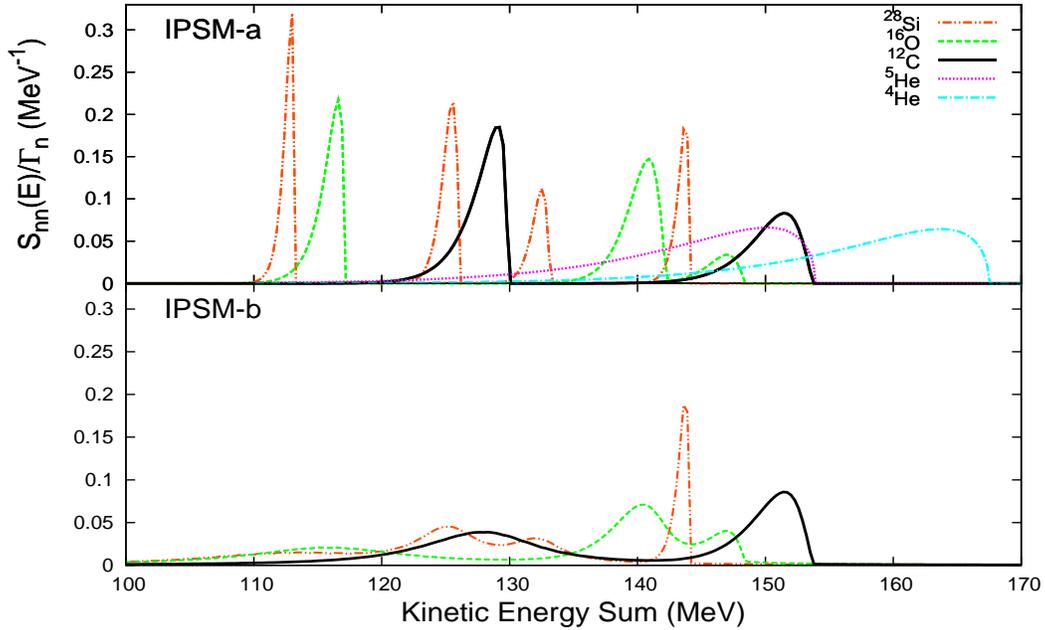}
\end{center}
\caption{\footnotesize
(Color online) Normalized energy spectra  $ S_{nn}(E)/\Gamma_n$ for
$^{4}_\Lambda$He, $^{5}_\Lambda$He, $^{12}_\Lambda$C, $^{16}_\Lambda$O, and
$^{28}_\Lambda$Si hypernuclei for the full OMEP, obtained within the approaches 
IPSM-a (upper panel) and  IPSM-b (lower panel). 
For the $s$-shell hypernuclei, only the IPSM-a approach has been used.}
\label{Fig2}\end{figure}

In Figs. \ref{Fig1} and \ref{Fig2}  we show, respectively, the
normalized energy spectra  $ S_{np}(E)/\Gamma_p$ and $
S_{nn}(E)/\Gamma_n$ for $^{4}_\Lambda$He, $^{5}_\Lambda$He,
$^{12}_\Lambda$C, $^{16}_\Lambda$O, and $^{28}_\Lambda$Si
hypernuclei, evaluated within the full OMEP, that comprises the
($\pi,\eta,K,\rho,\omega,K^*$) mesons. 
The single-particle energies for the strictly stationary hole states 
have been taken from 
Wapstra and Gove's compilation~\cite{Wa71}, and those of the quasi-stationary 
ones have been estimated from the studies of the quasi-free scattering 
processes $(p,2p)$ and  $(e,e'p)$
\cite{Ja73,Fr84,Be85,Le94,Ya96,Ya01,Yo03,Ya04,Ko06}.

The two IPSM approaches exhibit some quite important differences:
\bit
 \item[a)] {\em IPSM-a}:  The spectra cover the
energy region $110$ MeV $<E< 170$ MeV and contain one or more
peaks, the number of which is equal to the number of shell-model orbitals
$ 1s_{1/2}, 1p_{3/2}, 1p_{1/2}, 1d_{5/2}, 2s_{1/2}, 1d_{3/2} \cdots$
that are either fully or partly occupied in $\ket{J_C}$.
Before including the recoil, all these peaks would be just spikes
at the liberated energies $\Delta_{j_N}$, as can be seen from \rf{3} setting 
$E_r=0$. With the recoil effect, they behave as 
\be
S_{nN}(E \cong\Delta_{j_N})\sim \sqrt{(\Delta_{j_N}-E)(E-\Delta'_{j_N})}
e^{-M(A-2)(\Delta_{j_N}-E)b^2},
\label{35}
\ee
and develop rather narrow widths $\sim [b^2M(A-2)]^{-1}$, where  $b$ is
the harmonic oscillator  size parameter, which  has been  taken from 
Ref.~\cite{It02}.
These  widths go  from $\cong 3$ MeV for $^{28}_\Lambda$Si to $\cong 20$ MeV 
for $^{4}_\Lambda$He, as indicated in the upper panels of the just mentioned 
figures.
\item[b)]{\em IPSM-b}:
In the lower panels of the same figures are shown the results obtained
when the recoil is  convoluted with  the  Breit-Wigner distributions
\rf{34} for the strength functions of the fragmented deep hole states.
The widths $\gamma_{j_N}$ have been estimated from 
Refs.~\cite{Ja73,Fr84,Be85,Ma85, Le94,Ya96,Ya01,Yo03,Ya04,Ko06}, 
and in particular from~\cite[Figure 11]{Ja73} and~\cite[Table 1]{Ya96}, 
with the results:
$\gamma_{{1s_{1/2}}}=9$ MeV in $^{12}_\Lambda$C,
$\gamma_{{1s_{1/2}}}=14$ MeV 
and $\gamma_{{1p_{3/2}}}=3$ MeV in $^{16}_\Lambda$O 
\footnote{The  ${3/2}_1^-$ peak is at  $6.32$ MeV, but  small amounts of
the $p_{3/2}$ strength are also fragmented to the states of
$9.93$ MeV and $10.7$ MeV~\cite{Ko06}.}, and
$\gamma_{{1s_{1/2}}}=16$ MeV 
and $\gamma_{{1p_{3/2}}}=\gamma_{{1p_{1/2}}}=5$ MeV in $^{28}_\Lambda$Si,
both for protons and neutrons.
One sees that, except for the ground states, the narrow peaks engendered by the 
recoil effect become now pretty  wide  bumps.
\eit

We feel that the above  rather rudimentary  parameterization  could be
realistic enough for a qualitative discussion of the kinetic energy sum spectra.
A more accurate  model should be probably  necessary for  a full quantitative 
study and comparison with data.

\section{ General considerations and connection with data \label{Gen}}

\begin{figure}[h]
\begin{center}
   \leavevmode
   \epsfxsize = 14cm
     \epsfysize = 9cm
\epsffile{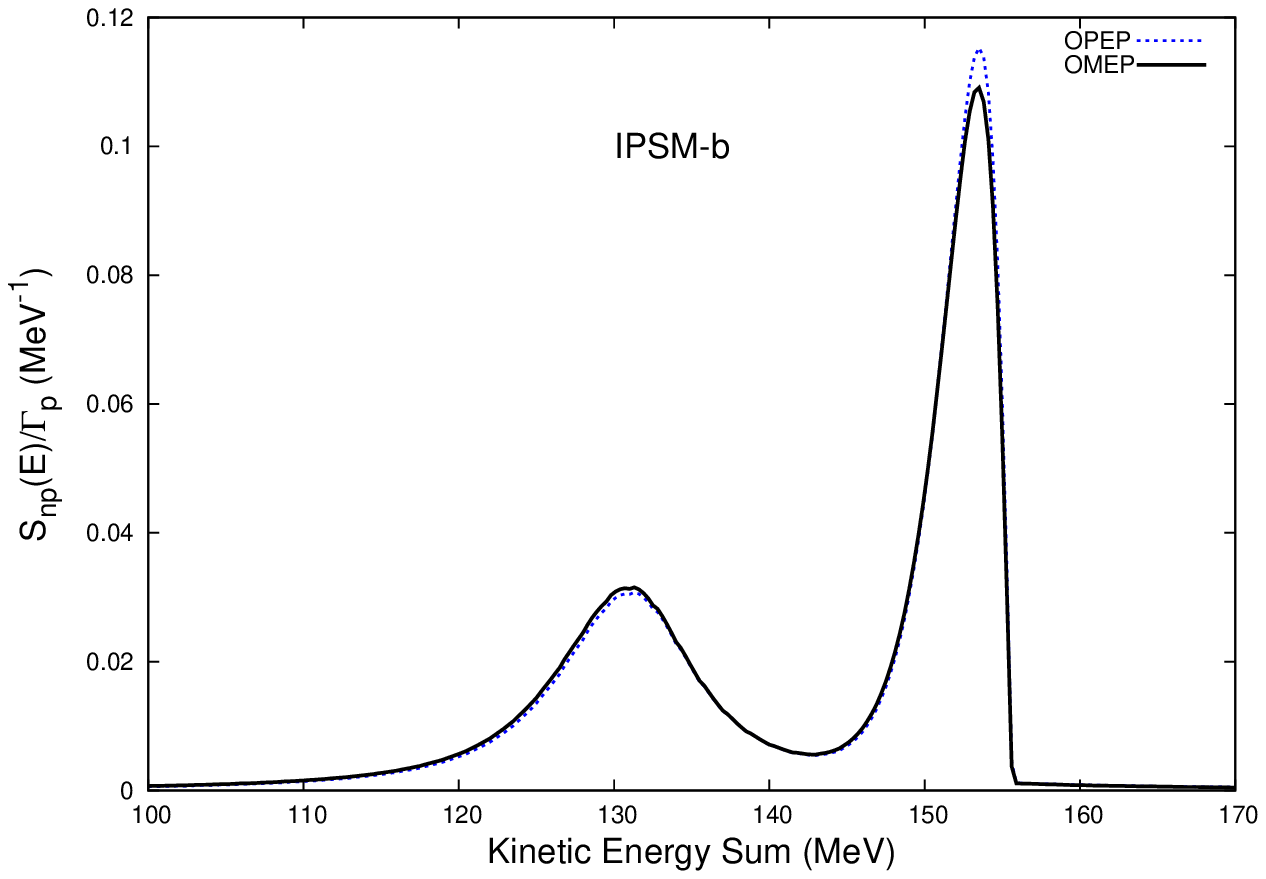}
\end{center}
\caption{\footnotesize
(Color online) Normalized energy spectra  $ S_{np}(E)/\Gamma_p$ for the decay
of $^{12}_\Lambda$C, computed with the full OMEP and the OPEP in the framework 
of the IPSM-b.}
\label{Fig3}\end{figure}

The normalized spectra shown in Figs. \ref{Fig1} and \ref{Fig2} have a 
very weak dependence on the dynamics involved in the NMWD process proper, 
and almost identical shapes would have been obtained if only the 
One-Pion Exchange Potential (OPEP) had been taken into account.
To understand this fact one can appeal to the  s-wave approximation, 
which assumes that only the relative matrix elements of the
form $\bra{p,lSJT}V\ket{0JJT}$, \ie with the $\Lambda N$ system in an s-state,
significantly contribute to
$\M(pPlL\lambda SJT;j_N)$. This has been examined quantitatively, for $p$-shell
hypernuclei, in Ref.~\cite{Be92}; see also the Refs.~\cite{Ba02,It02}.
 Furthermore,  as we have discussed
in Ref. \cite{Ba07}, those matrix elements depend only very weakly on the
relative momentum $p$ and, as such,  they  can be  evaluated  at  the maximum
value of $p=p_\Delta=\sqrt{M\Delta}$, which corresponds to $P=0$ or, according
to Eq. \rf{19}, to $E=\Delta_{j_N} \cong \Delta$. Thus the energy dependence of
$S_{nN}(E)/\Gamma_N$ remains exclusively in kinematical  phase-space factors,
and the position and width of the peaks will be unaffected by the dynamics of
the decay process, which will influence, to some extent, only their relative
heights. This is illustrated in the case of the $np$ spectrum for
$^{12}_\Lambda$C in Fig. \ref{Fig3}, showing that even this latter effect is
very small.
The comparison between the normalized spectra obtained with the full OMEP and 
with the $\pi+K$ exchange potential would give an almost perfect superposition.

Needless to stress that the transition probabilities $\Gamma_N$ do strongly
depend on the hypernuclear transition potential, but this dependence is washed
out in the ratios defining the normalized spectra. Conversely,
for a given choice of transition potential 
both shell model approaches discussed here yield very similar results 
for the $\Gamma_N$. These points are illustrated in Table \ref{t1}, 
where one can also see that, as already anticipated in Section \ref{Kin}, 
the effect of the recoil on these quantities is negligible.

\begin{table}
\caption{\label{t1} Nonmesonic decay rates in units of 
$\Gamma_\Lambda^{0}=2.50 \times 10^{-6}$ eV and $n/p$ branching ratios for 
$^{12}_\Lambda$C, $^{16}_\Lambda$O and $^{28}_\Lambda$Si computed with several 
transition potentials and using the \mbox{IPSM-a} and \mbox{IPSM-b} approaches. 
The values obtained in the \mbox{IPSM-a} framework but neglecting the recoil 
are shown within parentheses.}
\begin{ruledtabular}
\begin{tabular}{lccc}
                        & $\Gamma_n$ & $\Gamma_p$ & $\Gamma_n/\Gamma_p$  \\ 
\hline
$^{12}_\Lambda$C        &            &            &                      \\
IPSM-a & & & \\
OMEP               & 0.249 (0.249) & 0.956 (0.960) & 0.260 (0.259)       \\
$\pi+K$            & 0.244 (0.244) & 0.755 (0.758) & 0.323 (0.322)       \\
IPSM-b & & & \\
OMEP                    &  0.246     & 0.947      & 0.260                \\
$\pi+K$                 &  0.241     & 0.748      & 0.322                \\
OPEP                    &  0.142     & 1.004      & 0.141                \\
\hline
$^{16}_\Lambda$O        &            &            &                      \\
IPSM-a & & & \\
OMEP               & 0.290 (0.290) & 1.024 (1.027) & 0.283 (0.282)       \\
$\pi+K$            & 0.287 (0.287) & 0.811 (0.813) & 0.354 (0.353)       \\
IPSM-b & & & \\
OMEP                    & 0.285      & 1.009      & 0.282                \\
$\pi+K$                 & 0.282      & 0.799      & 0.353                \\
\hline
$^{28}_\Lambda$Si       &            &            &                      \\
IPSM-a & & & \\
OMEP               & 0.348 (0.348) & 1.163 (1.164) & 0.299 (0.299)       \\
$\pi+K$            & 0.341 (0.341) & 0.934 (0.935) & 0.365 (0.365)       \\
IPSM-b & & & \\
OMEP                    & 0.345      & 1.123      & 0.307                \\
$\pi+K$                 & 0.338      & 0.903      & 0.374                \\
\end{tabular}
\end{ruledtabular}
\end{table}

One can then summarize our findings by saying  that in light systems
($^{4}_\Lambda$He and $^{5}_\Lambda$He)
the kinetic energy sum  coincidence spectra
$S_{nN}(E)$, normalized to the total decay rates $\Gamma_{N}$,
basically depend on energies  associated with the three-body kinematics.
The differences between $S_{np}(E)$ and $S_{nn}(E)$  are
mainly due to the differences in the proton and neutron separation energies
and in the spectroscopic factors  $F^J_{j_N}$. For the remaining hypernuclei it
is imperative, in addition, to take into account that  most of the 
hole states are fragmented and consequently one has to consider the spreading
of their strengths.

The residual interaction among the valence particles
and their coupling to the collective rotational and/or vibrational 
motions  are not explicitly considered in the present work. 
But it is not expected that these would modify qualitatively the above scenario.
It might be worth noting, nevertheless, that the  pairing force would be 
capable of shifting some of the strength from the occupied levels  to higher
lying orbitals. For instance, in $^{12}_\Lambda$C a part of the
$1p_{3/2}$ strength would be moved up into the $1p_{1/2}$ orbital,
while in  $^{28}_\Lambda$Si the  strength would be moved from the $1d_{5/2}$
level into the empty $ 2s_{1/2}$ and $  1d_{3/2}$  states.
However, we believe that the coupling of the deep-hole states to
other more complicated configurations through the residual interaction, 
which is treated here in a phenomenological way,
is by far a more  relevant effect for the physics
discussed in the present paper.

We would not like end this paper without making some comments on the relation 
between the formalism developed here and the experiments.
To this end we shall follow closely the discussion in Ref.~\cite{Bau08}.
The theoretical prediction for the number of $nN$ pairs detected in coincidence 
with kinetic energy sum $E$ within the interval $dE$ can be written as
 \be
d{\rm N}_{nN}(E)=C_{nN}(E)S_{nN}(E)dE,
\label{36}\ee
where the factor $C_{nN}(E)$ depends on the experimental environment
and includes all quantities and effects not considered in $S_{nN}(E)$, such as
the  number of produced hypernuclei, the detection efficiency and acceptance, 
\etc. 
Assuming, for simplicity, that the experimental spectra have already been 
corrected for detection efficiency and acceptance and that the possible 
remaining energy dependence in this factor can be neglected, the predicted 
total number of detected events ${\rm N}_{nN}$ can be related to $\Gamma_N$ 
as follows:
 \be
{\rm N}_{nN}=\int \frac{d{\rm N}_{nN}(E)}{dE}dE=C_{nN}
\int S_{nN}(E)dE=C_{nN}\Gamma_N.
\label{37}\ee
This allows us to rewrite \rf{36} in the form
 \be
\frac{d{\rm N}_{nN}(E)}{dE}={{\rm N}_{nN}}\frac{S_{nN}(E)}{\Gamma_N}.
\label{38}\ee
What is measured in an experiment is the number of  pairs 
$\Delta {\rm N}_{nN}^{exp}(E_i)$  at a given energy $ E_i$
within a fixed energy bin $\Delta E_{nN}$, \ie 
$\Delta {\rm N}_{nN}^{exp}(E_i)/\Delta E_{nN}$.
The total number of observed events is
\be
{\rm N}_{nN}^{exp}=\sum_{i=1}^{m} \Delta {\rm N}_{nN}^{exp}(E_i),
\label{39}\ee
where $m$ is the number of bins.
The  spectrum  $S_{nN}(E)$ can be normalized to the experimental one
by  identifying ${\rm N}_{nN}$ in \rf{38} with ${\rm N}_{nN}^{exp}$.
Thus, the quantity that we have to confront with measurements is
 \be
\Delta {\rm N}_{nN}(E)= {\rm N}_{nN}^{exp}\Delta E_{nN}\frac{S_N(E)}{\Gamma_N}.
 \label{40}\ee
For instance, to compare  the experimental data given in ~\cite[Fig.11 ]{Pa07} 
with our calculations shown in  Figs. \ref{Fig1} and \ref{Fig2}, the latter 
should be multiplied by the factors 
${\rm N}_{np}^{exp}\cdot\Delta E_{np}=87\times 5$ MeV $=435$ MeV,
and ${\rm N}_{nn}^{exp}\cdot\Delta E_{nn}=19\times 5$ MeV $=95$ MeV, 
respectively.

Here we have, neither considered the resolution of the detector system,
nor explicitly included the FSI. Also, we have ignored the three-body
$\Lambda NN$ decay contributions.
As a consequence it is very reasonable that only a wide bump at about $140$ MeV
would appear in the experimental spectra for $^{12}_\Lambda$C and heavier
hypernuclei. 
The IPSM predicts quite similar spectra for the $np$ and $nn$ pairs.
Moreover, from the present results one could venture to say that the neutron
bump should lie at a smaller energy than the proton one. This  agrees only
marginally with the experiments performed so far, where important differences
between the $np$ and $nn$ spectra have been observed.

\section{Concluding remarks \label{Con}}

In this paper, we have investigated, in the framework of the independent 
particle shell model (IPSM), the effects of the recoil of the residual 
nucleus and of the spreading in strength of the deep-hole states on nonmesonic 
weak decay (NMWD) observables. We conclude that, while their effect 
is of minor importance for integrated observables like the decay rates and the 
$n/p$ branching ratio, they play a crucial role in determining the 
shapes of the normalized kinetic energy sum coincidence spectra of $nn$ and $np$
pairs. For the spectra of $s$-shell hypernuclei, the recoil effect is the 
most important one.

In summary, we believe that  the IPSM is the appropriate lowest-order 
approximation for the theoretical calculation of the two-particle spectra in 
the NMWD when: 1) the recoil effect is included, 
and 2) the fragmentation  of the strengths of the deep-hole states
is taken into account. It is in comparison to this picture that one
should appraise the effects of the  FSI and of the two-nucleon-induced 
decay mode.

The consequences of the two effects dealt with here on the one-particle 
kinetic 
energy spectra and on the opening angle distributions of $np$ and $nn$  pairs 
will be discussed elsewhere.

\begin{acknowledgments}
This work was partly supported by the Brazilian agencies CNPq and FAPESP, and 
by the Argentinian agency CONICET under contract PIP 6159. 
M. S. Hussein thanks the Martin Gutzwiller program at the 
Max Planck Institute for the Physics of Complex Systems-Dresden
for support.

\end{acknowledgments}

\end{document}